# Near-bandgap wavelength-dependent studies of long-lived traveling coherent longitudinal acoustic phonon oscillations in GaSb/GaAs systems

J. K. Miller,[1] J. Qi,[1] Y. Xu,[1] Y.-J. Cho,[2] X. Liu,[2] J. K. Furdyna,[2]
I. Perakis,[3] T. V. Shahbazyan,[4] and N. Tolk[1]

[1] Department of Physics and Astronomy, Vanderbilt University, Nashville, TN, 37235
[2] Department of Physics, University of Notre Dame, Notre Dame, IN 46556
[3] Department of Physics, University of Crete, 71003, Greece
[4] Department of Physics, Jackson State University, MS 39217

We report first studies of long-lived oscillations in femtosecond optical pump-probe measurements on GaSb/GaAs systems. The oscillations arise from a photo-generated coherent longitudinal acoustic phonon wave, which travels from the top surface of GaSb across the interface into the GaAs substrate, thus providing information on the optical properties of the material as a function of time/depth. Wavelength-dependent studies of the oscillations near the bandgap of GaAs indicate strong correlations to the optical properties of GaAs.

Coherent phonon oscillations generated by ultrafast optical excitation has been observed in various material systems using time resolved pump-probe schemes [1-5]. In these experiments, femtosecond pump light pulses are absorbed at sample surfaces, resulting in the generation of coherent phonons. The time delayed probe pulses sense the coherent phonon modulation of optical properties, such as reflectance or transmittance. In particular, traveling coherent longitudinal acoustic phonon (CLAP) waves have been recently reported in a wide range of materials, including manganites [6], and III-V semiconductors [7-9]. Here, the photo-induced acoustic phonon wave generated near the surface travels into the substrate following excitation. The oscillations in the pump-probe reflectance responses arise from interference between probe laser photons reflected from the front surface and from the traveling coherent longitudinal *acoustic* phonon plane wave. This is in contrast to the observed short-lived, fast oscillations arising from the coherent *optical* phonon modulation of the optical properties of the near surface.

We report here first observations of long-lived traveling coherent longitudinal acoustic phonon (CLAP) waves in the GaSb/GaAs system. The oscillatory response is strongly dependent on the probe light wavelength near the band gap of GaAs and persists for the entire measuring time window and beyond. Since the traveling CLAP wave differentially samples the material as it proceeds from material to material across interfaces, it constitutes a promising non-invasive technique for measuring optical properties of the material as a function of depth. It is also a useful tool for measuring layer thicknesses and band gap energies.

In the present study, we investigated GaSb/GaAs heterostructures, grown by standard molecular beam epitaxy (MBE). First, a semi-insulating "epi-ready" (001) GaAs substrate is heated to 600°C under $As_2$ flux for deoxidization. After this thermal cleaning, a 100nm GaAs buffer layer is grown on substrates at normal GaAs growth conditions (~600°C) in order to obtain an atomically flat surface. Finally, a layer of GaSb is grown at 490°C.

Standard femtosecond, time-resolved optical pump-probe measurements of transient reflectivity change $\Delta R/R$ are performed on these samples. The light source is a Coherent MIRA 900 Ti:Sapphire laser, which produced 150-fs-wide pulses in the 740 nm (1.675 eV) to 890 nm (1.393 eV) wavelength range at a repetition rate of 76 MHz. The oscillatory response can be observed at all temperatures from 4 K up to room temperature, and we did not observe strong temperature dependence of the oscillatory signal. However, the bandgap of GaAs shifts with temperature. Here we choose 30 K for our systematic wavelength dependent study, since the bandgap of GaAs at 30 K (1.517 eV or 817 nm in optical wavelength) is in the middle of our Ti:Sapphire laser's wavelength tuning range, while the room temperature bandgap of GaAs is very close to the upper limit of the tuning range (1.424 eV or 870 nm). The pump and probe beams are cross-polarized and their spot sizes on the sample are about 100 $\mu$m in diameter, with their energy ratio of 10:1. Typical pump light has an average power of 30 mW, equivalent to a pulse energy of 0.4 nJ and a fluence of 5.1 $\mu J/cm^2$. This excitation intensity is about 3 orders of magnitude smaller than the intensity in some references, where an OPA system is typically used. The pump beam is incident normal on the sample while the probe is at an angle of 45°.

In Figure 1, we show the measured pump-probe response at room temperature on a sample with a relatively thin, 20 nm GaSb layer, on a GaAs substrate. The wavelength is 890 nm (1.393 eV), slightly below the bandgap of GaAs (1.424 eV) at room temperature, but above the bandgap of GaSb (0.8 eV). It is clear that the total response consists of a fast transient (on the order of a few picoseconds) followed by a very long-lasting damped oscillatory tail. The initial fast transient is typical for carrier dynamics due to the excitation of hot photoexcited electron-hole pairs and relaxation by means of electron-phonon interaction in GaSb. The long-lasting oscillations are not directly related to the pump light since it only appeared after the first fast transient. The inset is the measured oscillatory response (open diamond), which is obtained by subtracting the thermal relaxation background, fitted with an exponentially damped oscillation function (solid line). Applying the fitting curve to the data, we obtain a period of oscillation of about 26.6 ps and the damping time of about 720 ps.

The origin of this class of oscillations was explained first by Thomsen *et al.* in Ref. [1]. Namely the ultrafast pump pulse is absorbed at the surface of the film and the absorbed photon energy gives rise to a transient electron and phonon temperature increase within the illuminated area, which then sets up a transient stress at the sample surface. This stress induces a strain wave (coherent longitudinal acoustic phonons), which propagates away from the surface at the speed of the LA phonons. This CLAP wave modifies the local dielectric constants and creates a discontinuity. When the probe pulse is incident onto the sample, part of the light is reflected from the top surface of the sample, and the rest of the transmitted light reaches the CLAP wave inside and gets partially reflected due to the discontinuity in the dielectric constants. As the CLAP wave travels, the top sample surface and the strain wave surface act as an interferometer. Oscillations arise from the fact that the distance between the two surfaces is constantly changing, causing a periodic phase difference between the two reflected beams [10].

In the GaSb/GaAs system, the top layer of GaSb is an absorber to create the phonon wave after photon excitation. The bandgap of GaSb is 0.8 eV, much smaller than the pump photon energy within the whole wavelength tuning range. On the other hand, the GaAs substrate has a bandgap very close to the pump photon energy. The absorption coefficient is about 5 times bigger in GaSb compared to GaAs at 1.5 eV [11]. Thus we assume that the pump photon energy is absorbed in the GaSb layer only. A temperature increase in GaSb due to excitation happens after the initial excitation of electrons and holes and electron-phonon interaction. These processes take place in several picoseconds. From the model proposed by Thomsen *et al.*, it is clear that the wavelength of the pump light plays little role in creating phonon waves, as long as the excitation photon energy can be absorbed by the top layer of GaSb. The thickness of the GaSb layer is small (20 nm), so that we assume the whole GaSb layer is excited to generate the CLAP wave. The CLAP wave travels along the normal direction at the speed of the LA phonons. When the probe pulse reaches the sample after excitation, part of the light is reflected from the air/GaSb layer, and part of the light penetrates into the GaSb and eventually into the GaAs. The probe light is then reflected back from the CLAP wave and produces interference with the reflected light from the air/GaSb surface. The oscillation period T can be calculated to be,

$$T = \lambda/(2n V_s \cos\theta), \tag{1}$$

where $\lambda$ is the probe wavelength, n is the index of refraction of the material, $V_s$ is the speed of LA phonons, and $\theta$ is the incident angle of probe light in GaAs. For the probe wavelength of 890 nm, as shown in Figure 1, we use values from the literature: n = 3.65 [11], $V_s = 4.73 \times 10^3$ m/s [12] and an incident angle of 45 degrees of the probe light in the air. The calculated period is 26.3 ps, in close agreement with the measured period of 26.6 ps.

We then perform wavelength dependent studies of the oscillations near the bandgap of

GaAs and the oscillation data are illustrated in Figure 2. In both Fig. 1 and Fig. 2, the oscillations are seen to persist with little damping for at least 1 ns, for photon energies below the bandgap. Above the bandgap, this decay is not due to the decay of the phonon wave itself, but rather to the above-bandgap optical absorption of GaAs. Thus, we conclude that the CLAP wave, which does not depend on the probe wavelength, remains coherent without significant decay for at least 1 ns for all probe wavelengths. This is also in good agreement with the 'near-field' assumption in [1]. Similar long-lasting CLAP oscillations have been previously observed [4, 8].

We may use a simple model to consider the experimental data shown in Fig. 2. The photoexcited carriers in the GaSb layer transfer their excess energy, deposited by the pump pulse, to LA phonons thus causing rapid expansion of the lattice on a subpicosecond timescale. The propagating CLAP wave can be viewed as a strained layer moving away from the surface with LA phonon velocity Vs. The measured signal is the result of interference between the probe beams reflected from the air/bulk interface and the bulk/strained layer boundary at the distance $z(t)=V_s t$ from the surface. The reflection coefficient in this three layer system is

$$R = \left| \frac{r_{12} + r_{23} e^{2ikz}}{1 + r_{12} r_{23} e^{2ikz}} \right|^2, \qquad (2)$$

where $r_{ij}=(N_i-N_j)/(N_i+N_j)$ is the Fresnel reflection amplitude from the boundary between media $i$ and $j$, and $N_i$ is complex refraction coefficient in media $i$. Here $N_1=1$ in air, $N_2=N(\omega)=n(\omega)+i\kappa(\omega)$ in the bulk semiconductor, $N_3=N(\omega)+\delta N(\omega)$ in the strained layer, and $k=(\omega/c)N\cos\theta$ is the normal component of the wavevector in the bulk. Thus we define $r_{12}$ as the reflection amplitude from the air/GaSb boundary, and $r_{23} \sim \delta N/N$ describes the reflection from the moving strained layer. Since the strain-induced change in refraction index, $\delta N$, is small, a simple expression for $R$ can be obtained by expanding Eq. (2) in the first order in small parameter $r_{23}$. The differential reflection, $\Delta R/R_0=(R-R_0)/R_0$, with $R_0=|r_{12}|^2$, then takes the form

$$\frac{\Delta R}{R_0} \approx A e^{-t/\tau} \sin\left(\frac{2\pi}{T} t + \varphi\right), \qquad (3)$$

where $A \propto \delta N/N$ is the amplitude, and $\varphi$ is the phase. The period $T$ is given by Eq. (1) and the damping time is defined as,

$$\tau = \lambda/(4\pi\kappa V_s \cos\theta) = Tn/2\pi\kappa. \qquad (4)$$

A more detailed model can be found in [1], which also leads to Eq. (3).

Figure 3 shows the period $T$, the damping time $\tau$, the amplitude A, and the initial phase $\varphi$, as a function of the wavelength as numerical fits to the measured oscillation data using Eq. (3). Figure 3 (a) shows that the oscillation period is close to linear with the probe light wavelength, in agreement with Eq. (1). Small variation from linearity may due to the fact that index of refraction of GaAs is not constant at different wavelengths near the bandgap. This linearity of oscillation period is a strong indication that a CLAP wave is indeed the origin of the long-lasting oscillatory response we observed. Figure 3(b) shows the damping time $\tau$ at different wavelengths. As discussed, the damping time of the oscillations reflects the optical absorption of GaAs, or the imaginary part of the refractive index $\kappa$ (from Eq. (4)). The onset wavelength of the increased damping time thus corresponds to the bandgap value from Fig. 3(b), suggesting a new way to determine the bandgap of GaAs. This figure also suggests that the traveling strain wave remains coherent for at least 1 ns, which corresponds to a traveling distance of about 5 μm. We expect the damping time in GaAs to be even longer with longer probe wavelengths. Figure 3(c) gives amplitude dependence. The amplitude of oscillations depends linearly on the strain-induced change in the refraction coefficient. The latter is determined by local reduction of the bandgap $\delta E_g$ due to the lattice expansion in the strained layer. The oscillations amplitude, $A \approx \delta N/N$, is then related to $\delta E_g$ as

$$A \propto \left| \frac{\partial N}{\partial \omega} \delta E_g \right|, \tag{5}$$

where the magnitude of $\delta E_g$ depends both on the intensity and frequency of pump beam. With increasing photon frequency, carriers with higher excess energy (with respect to GaSb bandgap) are created, leading to a larger energy transfer to the lattice. When the frequency is tuned away from the GaAs bandgap, this mechanism chiefly determines the oscillations amplitude. However, in the vicinity of the bandgap, $\omega \approx E_g$, the dielectric function of GaAs experiences a rapid change as the frequency passes through the absorption onset. Consequently, the derivative of $N$ in Eq. (5) exhibits a sharp peak near the bandgap, resulting in a strong peak in the amplitude $A$ in Fig. 3 (c). Figure 3(d) shows the initial phase $\varphi$ in Eq. (3) at different wavelengths. At wavelengths away from the bandgap, it represents the phase difference between the two reflected light beams when the strain wave first enters GaAs, which is then linear to the GaSb thickness and probe wavevector ($2\pi/\lambda$) [4]. Near the bandgap however, the phase experiences a jump as indicated by the sharp dip in the figure. This phase discontinuity is caused by the anomalous behavior of the phase shift at reflection from the strained layer that is determined by the Fresnel reflection coefficient $r_{23} \sim \delta N/N$.

All the above measurements and analysis have been carried out on a GaSb/GaAs sample with a 20 nm GaSb top layer. We have neglected the possibility of oscillations due to the CLAP wave traveling inside the GaSb layer because of the relatively small thickness. When the GaSb layer becomes thicker, the CLAP wave can produce oscillations while still in GaSb as well. Oscillations due to the CLAP wave in GaSb have been reported recently [7]. Our focus here is to observe how the pump induced CLAP wave propagates through two different materials, in our case GaSb and GaAs, which has not been studied previously. Figure 4 shows the measured oscillatory responses of GaSb/GaAs with a 500 nm thick GaSb top layer over a variety of wavelengths. The pump light still generates the traveling CLAP wave in GaSb. However, since the GaSb layer is thick, the generated CLAP wave is seen to travel through GaSb before it reaches the GaAs. The speed of LA phonons in GaSb at 30 K is $4.01 \times 10^3$ m/s [13]. It can be calculated that it takes about 120 ps for the strain wave to reach GaAs. At the GaSb/GaAs interface, most of the acoustic phonon wave enters GaAs; with less than 1% of the phonon energy reflected back [14]. We then ignore the reflected wave and assume the CLAP wave enters GaAs without loss.

The time that the phonon wave arrives at the interface is indicated by the red dash line in Fig. 4. This line separates the observed responses into two regimes. Before the red dash line, the CLAP wave propagates in the GaSb layer and produces oscillations as well. The amplitude of the oscillations damps relatively fast due to the absorption of probe light in GaSb. We have mentioned that the bandgap of GaSb is much smaller than the whole probe photon energy range, which accounts for the observed strong absorption. Responses in this regime are essentially wavelength independent. After the red dash line, the CLAP wave passes the GaSb/GaAs interface and continues to travel in GaAs layer. The observed responses resemble results in the thin GaSb/GaAs sample, with strong wavelength dependence similar to what have been demonstrated in Fig. 2. The lack of oscillations after the phonon wave enters GaAs at shorter wavelengths is due to strong absorption of both GaSb and GaAs. Although the two materials have different indices of refraction and speed of sound values, the oscillation periods calculated from Eq. (1) are very similar and cannot be distinguished from the experiment.

Interestingly, at the GaSb/GaAs interface, the oscillation amplitudes suddenly increase at wavelengths near the bandgap of GaAs in Fig. 4. The probe light intensity does not change abruptly across the interface. The intensity of the strain wave is material property related and does not have any probe light wavelength dependence. However, from the experimental results and above analysis on the first sample, we know that strain wave induced change of index of refraction, and hence the oscillation amplitude Eq. (5), has sharp wavelength dependence and peaks near the bandgap of GaAs. Thus, this sudden increase of oscillation amplitude in Figure 4 is associated with the wavelength dependent change of the dielectric constant in GaAs, compared to GaSb where the light wavelength is always far from the bandgap energy. It is worth pointing out that this sudden change of amplitude can be utilized to determine unknown film thickness, by simply calculating the propagating distance from the measured oscillatory responses, providing the speed of sound is known.

In summary, we have carried out a systematic wavelength dependent study of traveling CLAP waves in the GaSb/GaAs system. Oscillatory responses from time-resolved pump-probe reflectance measurements can be observed in both GaSb and GaAs and are attributed to the interference of different reflected probe beams due to a moving 'Fabry-Perot' interferometer configuration. The oscillation period is determined by the probe wavelength and by material properties, such as the index of refraction and the speed of LA phonons. We have focused our study mainly on GaAs, an important material both for research and application. This is the first detailed study of CLAP waves in GaAs. From our experimental results, the pump induced CLAP wave can last for at least 1 ns, and travels as deep as 5 μm. We have developed a model that describes the damping time and amplitude of the observed oscillations when the probe energy is tuned near the bandgap of GaAs. We have also analyzed situations when the CLAP wave travels through a materials interface. For GaSb/GaAs interface, the strain pulse remains essentially unchanged and passes the interface without significant reflection or dephasing. This technique holds the great promise to be utilized as a non-invasive tool to measure depth-dependent film thicknesses and optical properties. We gratefully acknowledge support from the Army Research Office and the National Science Foundation.

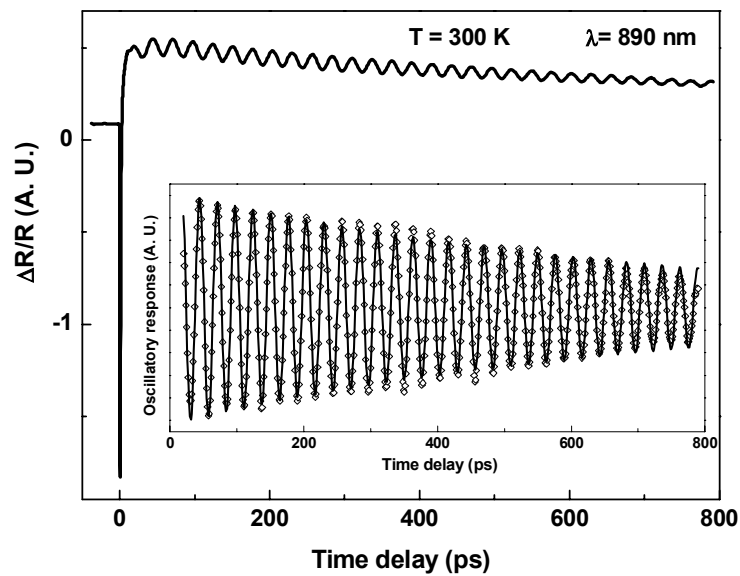

*Figure 1. Total pump-probe response of GaSb(20 nm)/GaAs at 890 nm at 300 K. Inset: the subtracted oscillatory response (open diamond) fitted with a damped oscillation simulation (solid line).*

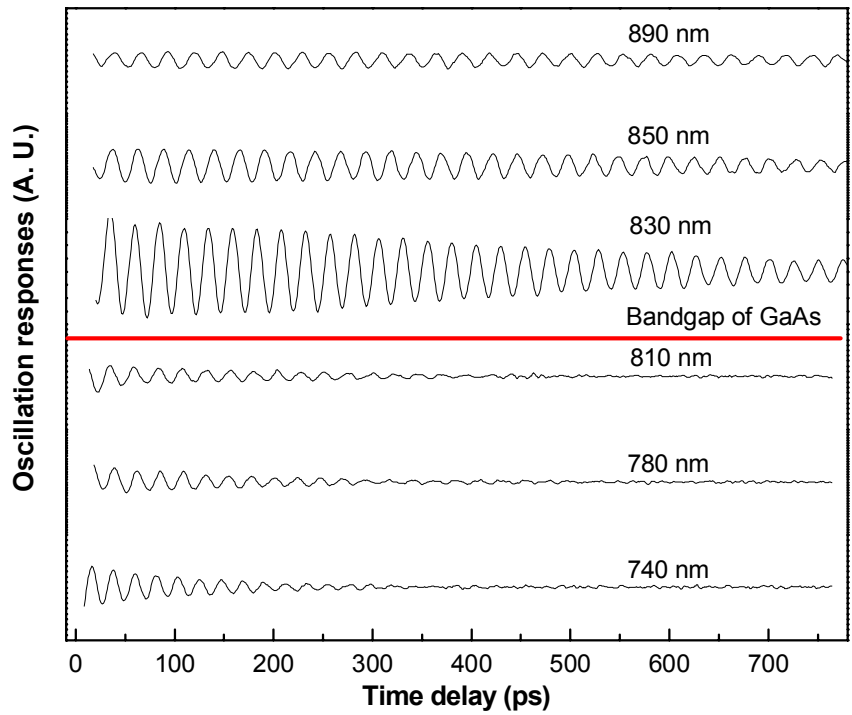

*Figure 2. Oscillatory responses measured at different wavelengths at 30 K.*

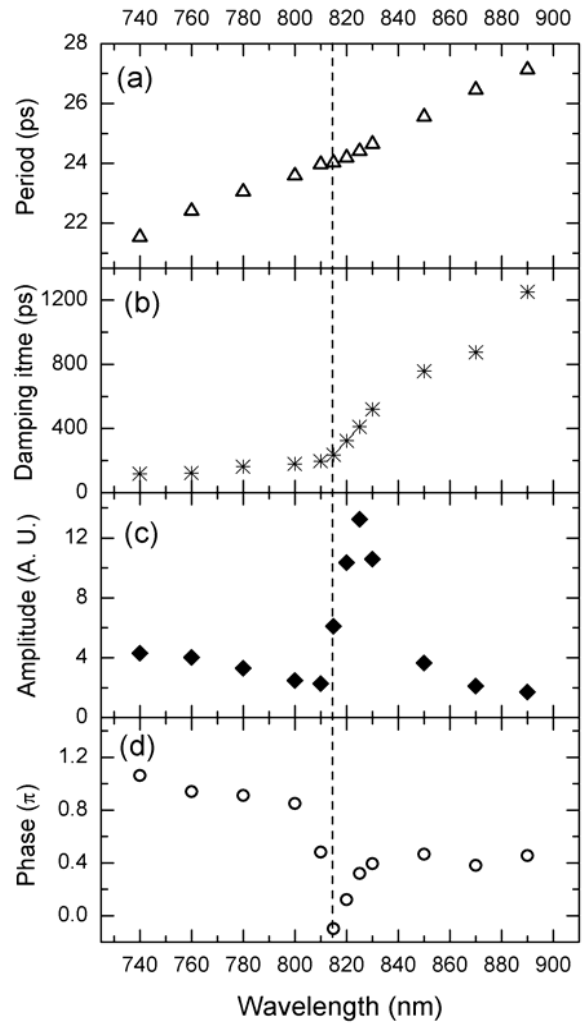

*Figure 3. Wavelength dependence of the oscillations parameters fitted with Eq. (3) at 30 K. The dash-dot line indicates the bandgap of GaAs at 30 K. (a) Oscillation period (open triangle) (b) Damping time (star) (c) Amplitude (solid diamond) (d) Initial phase (Open circle).*

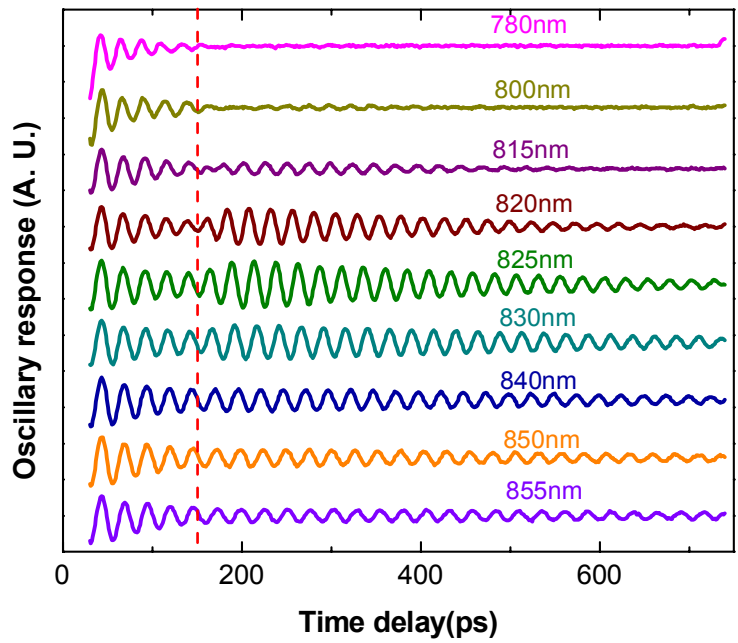

*Figure 4. Oscillatory responses at different wavelengths with 500 nm GaSb top layer. The dash line indicates when the strain wave travels from GaSb layer into the GaAs substrate.*

References:
1. C. Thomsen *et al.*, Phys. Rev. B 34, 4129 (1986).
2. T. Pfeifer *et al.*, Phys. Rev. Lett. 69, 3248 (1992)
3. C. Sun *et al.*, Phys. Rev. Lett. 84, 179 (2000)
4. Bozovic *et al.*, Phys. Rev. B 69, 132503 (2004)
5. M. Hase *et al.*, Phys. Rev. B 71, 184301 (2005)
6. D. Lim *et al.*, Phys. Rev. B 71, 134403 (2005)
7. J. Wang *et al.*, Phys. Rev. B 72, 153311 (2005)
8. S. Wu *et al.*, Appl. Phys. Lett. 88, 041917 (2006)
9. J. S. Yahng *et al.*, Phys. Lett. 80, 4723 (2002)
10. H.-N. Lin *et al.*,, J. Appl. Phys. 69, 3816 (1991)
11. D. E. Aspnes and A. A. Studna, Phys. Rev. B 27, 985 (1983).
12. J. S. Blakemore, J. Appl. Phys. 53, R123 (1982)
13. W. F. Boyle and R. J. Sladek, Phys. Rev. B 11, 2933 (1975)
14. A. Wood, in *Acoustics*, (Dover, New York, 1966)